# iQ Cavity for Material Permittivity Characterization

Kenneth W. Allen and Mark M. Scott

**Abstract.** *We present an X-band waveguide (WR90) and UHF waveguide (WR1500) measurement method that permits the extraction of the complex permittivity for low dielectric loss tangent material specimen. The extraction method relies on computational electromagnetic (CEM) simulations; coupled with a genetic algorithm; to fit the experimental measurement and the simulated transmitted scattering parameter ($S_{21}$) of the $TE_{10}$ mode through the waveguide with the material specimen partially filling the cross-section. This technique provides the material measurement community with the ability to accurately extract material properties of low-loss material specimen.*

**Introduction.**

Waveguide measurement techniques have been successfully implemented for the extraction of the complex permittivity ($\varepsilon(\omega)^* = \varepsilon(\omega)' - j\varepsilon(\omega)''$) of dielectric material specimen for nearly seven-decades [1-7]. However, more recently, there have been significant advances due to computational electromagnetic (CEM) simulation tools and global/local search algorithms [8,9]. The use of these computational tools has relaxed the geometrical restrictions, which allows for the characterization of complicated shapes of the specimen [9]. However, the most accurate methods of extracting the material properties of low loss dielectrics, at microwave frequencies, have been resonance methods [10-12]. Unfortunately, these resonant methods are inherently restricted to spot frequencies, if the extraction relies on perturbation theory and are difficult to model using CEM tools. Also, the presence of air gaps can be detrimental to the system accuracy. The method presented in this article enables broadband permittivity characterization of low dielectric loss tangent

(tan$\delta$) material specimen. Material properties are extracted through CEM-based analysis of the transmitted scattering parameters of an X-band and UHF waveguide with an effectively electrically long material specimen partially filling the cross-section of the waveguide. This is achieved by using electrically long specimen for the X-band waveguide and by multiple traverses through an electrically small specimen in the UHF waveguide, effectively increases the length, as illustrated in Figure 1. Inserting iris plates into the UHF waveguide and establishing an intermediate quality factor resonance generate the multiple paths through the sample.

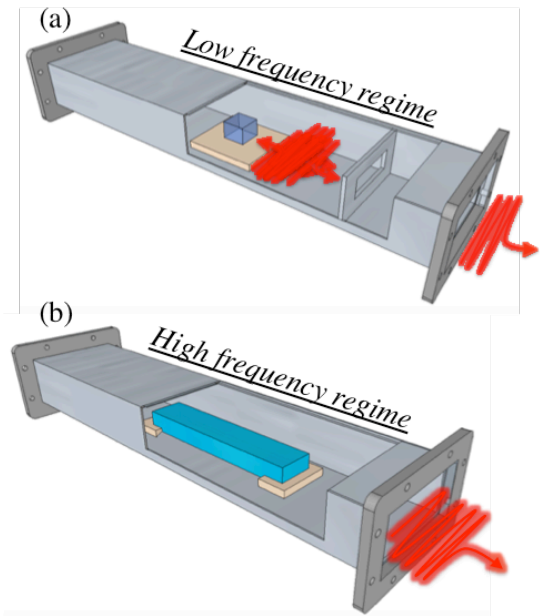

**Figure 1: Measurement systems for accurate low-loss tangent characterization for two different frequency regimes (a) low frequencies such as UHF and (b) high frequencies such as X-band.**

In general, complex permittivity extraction algorithm implements a finite element method (FEM) model that solves frequency-by-frequency the desired scattering parameters. Simulated s-parameters are then evaluated by a fitness function, which compares the simulated results to the experimentally obtained results. The result of the fitness function, in turn, drives a genetic algorithm that varies the complex permittivity values in the search space to match the simulated and experimental scattering parameters. This evaluation is performed over a range of frequencies, permitting the complex permittivity extraction over the entire band of interest. This method allows for the extraction of dispersive properties of the material specimen, which is not possible with high quality resonant methods.

**Technical Approach.**

These two primary methods can have advantageous depending on the frequency regime. In the case for high frequencies, such as X-band, a rectangular waveguide approach can be taken since the physical dimensions required to obtain an electrically long sample is not difficult (on the order of a few inches), as illustrated in Figure 2.

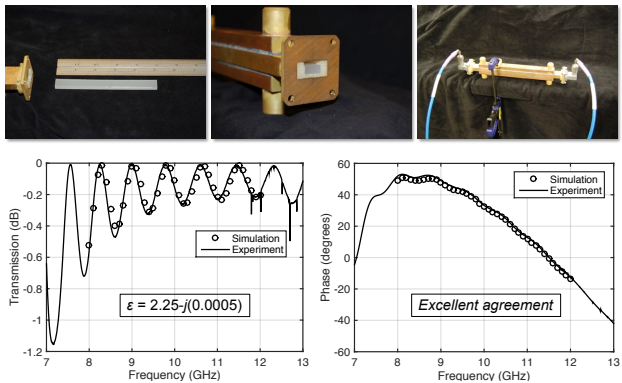

**Figure 2: Experimental measurement system (top row) and model-measurement plots of the transmitted scattering parameter for amplitude (bottom left) and phase (bottom right).**

The electrically long sample placed within the rectangular waveguide can support an intermediate-quality (iQ) factor resonance due to reflections between the front and back interface of the dielectric specimen.

This resonance can potentially lead to higher sensitivity to low loss tangent materials. In the case of low frequency characterization, it is difficult to obtain large volumes of the material in order to create an electrically long specimen; however, establishing an iQ factor within the waveguide can circumvent these issues. Placing iris plates in the rectangular waveguide will support multiple passes of the electromagnetic wave with the electrically small sample, in turn, increasing the interaction and potentially enhancing the sensitivity of the measurement. The quality factor of these resonances can be tuned by adjusting the dimensions of the iris plates within the UHF waveguide, as shown in Figure 3.

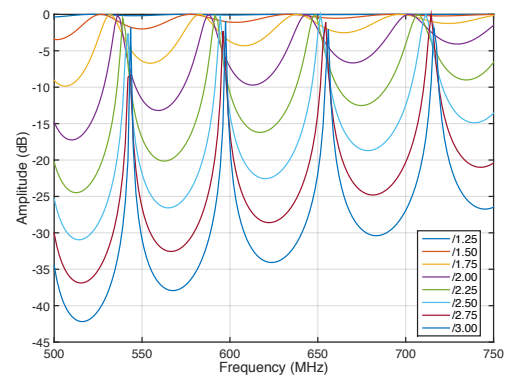

**Figure 3: Transmitted scattering parameter for amplitude from the UHF waveguide with iris plates tuned with the different fractions of the aspect ratio of the waveguides cross-section.**

**Technical Results.**

A new paradigm has emerged for measuring low loss material specimens as the result of investigating such iQ measurement techniques. These iQ techniques have illustrated, at the proof-of-concept level, that we can potentially have accurate broadband low loss tangent measurements. It has been shown that the sensitivity of the X-band waveguide measurements can be enhanced by increasing the electrical length of the sample. And likewise, increasing the quality factor of the resonance within the UHF waveguide can enhance the field interaction with the electrically small sample. We demonstrate the ability to tune the quality factor of the resonator by adjusting the iris plate dimension. This gives rise to an interplay between sensitivity enhancement and computational difficulty as the quality factor is increased; however, optimizing this tradeoff goes beyond the scope of this work.

A series of simulations were performed on two different software platforms in order to have code-code validation for the UHF waveguide measurement system, and these were also validated against experimental results, as shown in Figure 4.

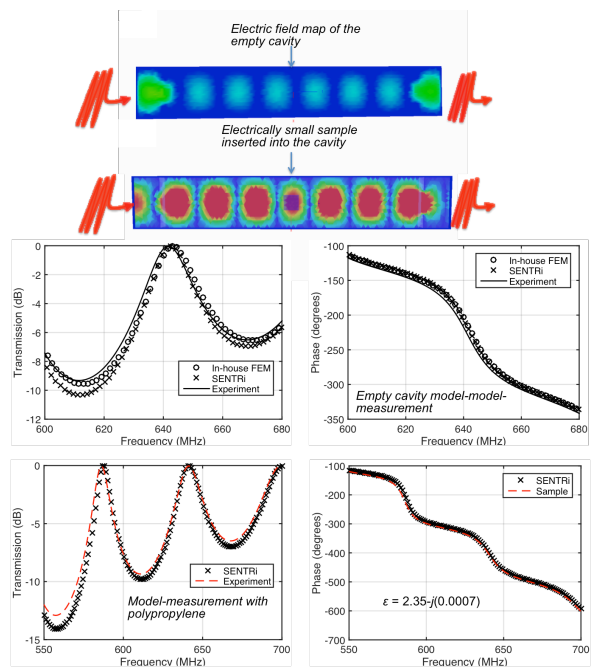

**Figure 4: Electric field maps within an empty cavity and a cavity with an electrically small specimen placed in the center. Two sets of scattering parameters for phase and amplitude of S$_{21}$ for the case of model-model-measurement agreement for an empty cavity and model-measurement for a cavity with an electrically small specimen.**

## Conclusions.

We illustrated, at the proof-of-concept level, measurement systems for low-loss dielectric material characterization for 2 separate frequency regimes rooted in intermediate-quality-factor resonances.

**Acknowledgements.** The author would like to thank GTRI for supporting this effort on IRAD.

**References.**

1. A. R. von Hippel, *Dielectric Materials and Applications*. Cambridge, MA, USA: MIT Press, 1954.
2. A. R. von Hipple, *Dielectrics and Waves*. New York, NY, USA: Wiley, 1945.
3. J.Baker-Jarvis, Transmission/reflection and short-circuit line permittivity measurements, NIST, Gaithersburg, MD, USA, Tech. Rep. 1341, 1995.
4. J.Baker-Jarvis, M. D. Janezic, J. H. Grosvenor, and R. G. Geyer, Transmission/reflection and short-circuit line methods for measuring permittivity and permeability, NIST, Gaithersburg, MD, USA, Tech. Rep. 1355, 1995.
5. W. B. Weir, Automatic measurement of complex dielectric constant and permeability at microwave frequencies, Proc. IEEE 62 (1974), 33-36.
6. M. M. Scott, D. L. Faircloth, J. A. Bean and K. W. Allen, Permittivity and permeability determination for high index specimens using partially filled shorted rectangular waveguides. Microw. Opt. Technol. Lett., 58: 1298–1301 (2016). doi: 10.1002/mop.29786
7. K. W. Allen, M. M. Scott, D. R. Reid, J. A. Bean, J. D. Ellis, A. P. Morris, and J. M. Marsh, "An X-Band Waveguide Measurement Technique for the Accurate Characterization of Materials with Low Dielectric Loss Permittivity," Rev. Sci. Instrum., vol. 87, issue 5 (2016).
8. M. E. Requena-Perez, A. Alberto-Ortiz, J. M. Monzo-Cabrera, and A. Diaz-Morcillo, Combined use of genetic algorithms and gradient descent methods for accurate inverse measurement, IEEE Trans. Microw. Theory Tech. 54 (2006), 615-624.
9. M. Scott, D. Faircloth, J. Bean, and S. Holliday, Biaxial permittivity and permeability determination for electrically-small material specimens of complex shape using shorted rectangular waveguides, IEEE Trans. Instrum. Meas. 63 (2014), 896-903.
10. B. Riddle, J. Baker-Jarvis, and J. Krupka, Complex permittivity measurements of common plastics over variable temperatures, IEEE Trans. Microw. Theory Tech. 51 (2003), 727-733.
11. J. Krupka, K. Derzakowski, B. Riddle, and J. Baker-Jarvis, A dielectric resonator for measurements of complex permittivity of low loss dielectric materials as a function of temperature, Meas. Sci. Technol. 9 (1998), 1751-1756.
12. E. J. Vanzura, R. G. Greyer, and M. D. Janezic, 60-millimeter diameter cylindrical cavity resonator: performance evaluation for permittivity measurements, Natl. Inst. Stand. Technol. Technical Note 1354, 1993.